\documentclass[twocolumn,showpacs,preprintnumbers,amsmath,amssymb]{revtex4}

\usepackage{graphicx}
\usepackage{dcolumn}
\usepackage{bm}

\begin{document}

\preprint{APS/123-QED}

\title{Quantum entanglement and drifting generated by an AC field resonant with frequency-doubled Bloch oscillations of correlated particles}

\author{W.S. Dias, F.A.B.F. de Moura, and M. L. Lyra}%
 
\affiliation{Instituto de F\'isica, Universidade Federal de Alagoas, 57072-900, Macei\'o-Alagoas, Brazil}


\date{\today}

\begin{abstract}

We show that initially localized and uncorrelated two-particles quantum wavepackets evolving in a one-dimensional discrete lattice become strongly entangled while drifting under the action of an harmonic AC field resonant with doubled Bloch oscillations promoted by a static DC field. Although partial entanglement is achieved when the AC field is resonant with the single-particle Bloch oscillations, it is strongly limited by the survival of anti-correlated unbounded states. We further show that the phase dependence of the wavepacket centroid velocity is similar to the semiclassical behavior depicted by a single-particle. However, the drift velocity exhibits a non-trivial non-monotonic dependence on the interaction strength, vanishing in the limit of uncorrelated particles, that unveils its competing influence on unbounded and bounded states.     

\end{abstract}

\pacs{03.67.Bg, 03.75.Lm, 37.10.Jk,, 67.85.-d}
\maketitle


Recent experiments investigating the behavior of few interacting atoms in optical lattices\cite{preiss} opened the possibility to generate a variety of quantum entangled matter states that can have a great impact on the search for universal and more efficient quantum computation processes, quantum sensing and metrology\cite{brennecke,childs,cronin}. It has been experimentally demonstrated that quantum entangled atom pairs perform Bloch oscillations (BOs) at twice the fundamental frequency in tilted optical lattices (TOLs)\cite{preiss}. BO is the phenomenon of oscillatory motion of  wavepackets placed in a periodic potential when driven by a constant force\cite{bloch,zener}. It has been  observed in several physical contexts such as semiconductors superlattices\cite{prlBOSEMI1,prlBOSEMI2}, ultra-cold atoms\cite{prlBOBEC1,arimondo,bonks,prlBOBEC2,prlBOBEC3}, optical\cite{prlBOSEMI2} and acoustic waves\cite{kosevich}. The coherent phenomenon of frequency doubling of BOs was predicted to occur in the presence of interaction\cite{wande1,wande2,wande3,longhi4,wande4} because two particles bind together and perform correlated tunneling. Fractional BOs at multiples of the fundamental BO frequency have also been demonstrated for larger clusters of interacting particles\cite{khomeriki}. A photonic realization of the  BOs frequency doubling has been proposed\cite{longhi3,longhi7} and experimentally achieved in waveguide lattices\cite{longhi5}. Its experimental demonstration with ultra-cold atoms of bosonic $^{87}$Rb in decoupled one-dimensional tubes of an optical lattice\cite{preiss} represents an important step towards the investigation of essential features of quantum many-body states.

Unidirectional transport and super-BOs can be induced when wavepackets are driven by superposed static and harmonic fields\cite{sbo1,sbo2,sbo3,kudo,sbo4,caetanolyra,sbo5}. It is achieved when the harmonic field is resonant with the underlying BO frequency, which depends linearly on the strength of the static field. A small detuning from the resonant condition results in super-BOs due to an effective tunneling renormalization. This phenomenology has been experimentally demonstrated in a weakly interacting Bose-Einstein condensate of Cs atoms placed in a TOL under forced
driving, achieving matter-wave
transport over macroscopic distances\cite{sbo3}. It has been theoretically demonstrated that some features of super-BOs and unidirectional transport require an important phase correction to be include in the tunneling renormalization picture\cite{sbo4}. Further, it has been theoretically shown that correlated super-BO of a bounded two-particles state can be observed under appropriate drive conditions\cite{sbo5}. Considering the current stage of experiments are in a position to probe the effect of interactions on driven transport of few correlated cold atoms\cite{sbo3}, it is of fundamental importance to have new schemes devised to generate and manipulate atomic entangled states in discrete lattices which can be used to probe quantum aspects of mater waves\cite{holthaus,mandel,entanglement}.

In this letter, we  show that two-interacting particles placed in a TOL can be coherently transported when an external harmonic field is made resonant with the doubled BO frequency. We also unveil the dependence of the wavepacket centroid velocity on the strength of the interaction as well on the relative phase of the harmonic field. Contrasting with the corresponding unidirectional transport for the resonant condition at the fundamental BO frequency, the degree of entanglement will be shown to continuously increase, with the particles developing positive spacial correlations due to the suppression of unbounded wavepacket components in the entire range of reachable drift velocities.

The dynamics of  two interacting particles placed in a linear discrete lattice of spacing $d$ under the action of superposed static DC and harmonic AC fields can be described in the framework of the tight-binding Hubbard model Hamiltonian as 
\begin{eqnarray} \label{hh}
H\hspace*{-0.25cm}&=&\hspace*{-0.25cm}\sum_{n,\sigma=1,2}\left[J(\hat{b}^{\dagger}_{n+1,\sigma}\hat{b}_{n,\sigma}+\hat{b}^{\dagger}_{n,\sigma}\hat{b}_{n+1,\sigma}) + eF(t)d n\hat{b}^{\dagger}_{n,\sigma}\hat{b}_{n,\sigma}\right] \nonumber \\ ~&+& \sum_n U\hat{b}^{\dagger}_{n,1}\hat{b}_{n,1}\hat{b}^{\dagger}_{n,2}\hat{b}_{n,2}
\label{hamiltonian}
\end{eqnarray}
where $\hat{b}_{n,\sigma}$ and $\hat{b}^{\dagger}_{n,\sigma}$ are the annihilation and
creation operators
for the particles of charge $e$ at site $n$ in spin state $\sigma$, $J$ is the hopping amplitude, and $U$ is the on-site Hubbard interaction. We are considering the particles distinguishable by their spin state. In what follows we will use units of $e=d=J=\hbar=1$. The applied field is assumed to be given by $F(t)=F_0+F_{\omega}\cos{(\omega t +\phi)}$ where $F_0$ stands for the amplitude DC field and $F_{\omega}$ the amplitude of the AC field with frequency $\omega$. In the absence of interaction, unidirectional transport occurs when the harmonic field is resonant with the fundamental BO frequency $\omega_B=F_0$ and its sub-multiples\cite{sbo4}. 
 
In order to follow the time evolution of the two-particles wavepacket, we solved the time dependent Schr\"odinger equation by expanding the wavevector in the Wannier representation $|\Psi(t)\rangle = \sum_{n_1,n_2} \psi_{n_1,n_2}(t)|n_1,n_2\rangle$, where the ket $|n_1,n_2\rangle$ represents a state with the particle in the spin state $\sigma=1$ at site $n_1$ and the other particle at site $n_2$.  The coupled motion equations for the wavepacket amplitudes
\begin{eqnarray}
\hspace*{-0.3cm}& &i\frac{d\psi_{n_1,n_2}(t)}{dt}=\psi_{n_{1}+1,n_{2}}(t)+\psi_{n_{1}-1,n_2}(t)+\psi_{n_{1},n_{2}+1}(t)\nonumber \\\hspace*{-0.2cm}&+&\psi_{n_{1},n_{2}-1}(t)+\Bigl[F(t)(n_1+n_2)+\delta_{n_1,n_2}U\Bigl]\psi_{n_{1},n_{2}}(t)
\label{recequation}
\end{eqnarray}
were solved numerically by using a high-order method  based on the Taylor expansion of the evolution operator $\Delta (\delta t  )=\exp{(-iH\delta t)} = 1 +\sum_{l=1}^{n_o}\frac{(-iH\delta t)^l}{l!}$, where $H$ is the two-particles  Hamiltonian.  Our calculations were taken by using $\delta t= 10^{-3}$ and the sum of  evolution operator truncated at $n_o=12$. The chain size was large enough to keep the wavefunction amplitude at the borders always smaller than $10^{-20}$. We followed the time-evolution of an initial wavepacket composed of a non-entangled Gaussian wavepacket
\begin{equation}\label{eq:gau}
\langle n_1,n_2|\Psi(t=0)\rangle=\frac{1}{A}e^{-[(n_1-n_{1}^{0})^2+(n_2-n_{2}^{0})^{2}]/4\sigma^2}
\end{equation}
centered at the initial position ($n_1^0, n_2^0$). $A$ is a normalization factor. In the following numerical results, we will consider initial wavepackets with $\Sigma=4\sigma^2 =10$ and $20$ and field strengths $F_0=0.6$ and $F_{\omega}=0.8F_0$\cite{sbo3,caetanolyra}. It is important to stress that no net displacement of the wavepacket centroid is achieved for an initial Fock state with the particles occupying a single site\cite{prlBOSEMI2}. In this case, the wavepacket dynamics is symmetric with respect to the initial position.

We start following the time evolution of the wavepacket centroid associated with each particle defined as
\begin{equation} \label{eq:nmedio}
\langle n_i(t)\rangle
=\sum_{n_1,n_2}(n_i)|\psi_{n_1,n_2}(t)|^2~~~~~,~~i=1,2 .
\end{equation}
Due to the symmetry of the initial state and interaction Hamiltonian, one has that $\langle n_1(t)\rangle = \langle n_2(t)\rangle$. In fig.~1 we plot the centroid evolution for both cases of the AC field resonant with the fundamental and doubled BO frequencies for an intermediate interaction strength. A net unidirectional modulated  motion of the wavepacket centroid is obtained, whose average velocity depends on the relative phase of the AC field.

\begin{figure}[!t]
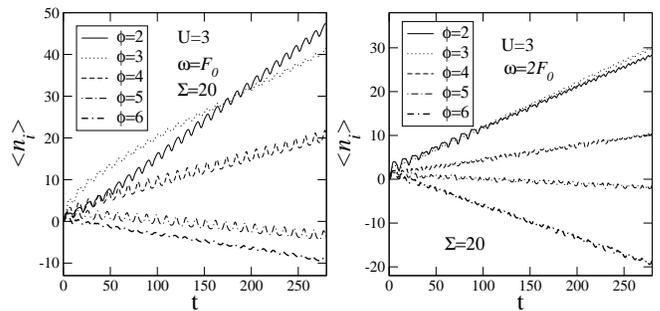

\begin{center}
\includegraphics*[clip,scale=0.27]{Fig1a.eps}
\includegraphics*[clip,scale=0.27]{Fig1b.eps}
\caption{ Time evolution of the one-particle wavepacket centroid. The initial state is composed by a direct product of Gaussian wavepackets with $\Sigma=20$ centered at $(n_1,n_2) = (0,0)$. The interaction strength is $U=3$. The AC and DC fields satisfy the resonance conditions $\omega=F_0$ (one-particle BO frequency) and $\omega=2F_0$ (correlated BO frequency). Results for distinct phases $\phi$ of the AC field are shown to stress the phase-dependence of the asymptotic centroid velocity.}
\label{fig1}
\end{center}
\end{figure}

As shown in fig.~1, a phase control of the AC field allows to tune both the speed and direction of the centroid motion. Within a semiclassical description, the wavepacket net velocity of a non-interacting particle driven by an AC field in resonance with the BO is given by \cite{caetanolyra}
\begin{equation}
v=2J_1(F_{\omega}/F_0)\cos{\left[(F_{\omega} /F_0)\cos{\phi}-\phi\right]}
\end{equation}
where $J_1(x)$ is the Bessel function of the first kind of order 1. In fig.~2, we plot the phase dependence of the average centroid velocity for the case of interacting particles driven by a resonant AC field. The phase dependence converges to the above semiclassical prediction as the initial wavepacket becomes wider, although presenting distinct overall amplitudes and reversed trends for the cases of resonance with the fundamental and doubled BO frequencies: larger (smaller) velocities are reached for wider initial wavepackets at the fundamental (doubled) BO resonance. No net transport occurs for $\phi=\pm\pi/2$ while maximum speeds are reached at $\phi_1=(F_{\omega}/F_0)\cos{\phi_1}\simeq 0.6411...$ (negative velocity) and $\phi_2=\pi-\phi_1=2.5008...$ (positive velocity).

\begin{figure}[!t]
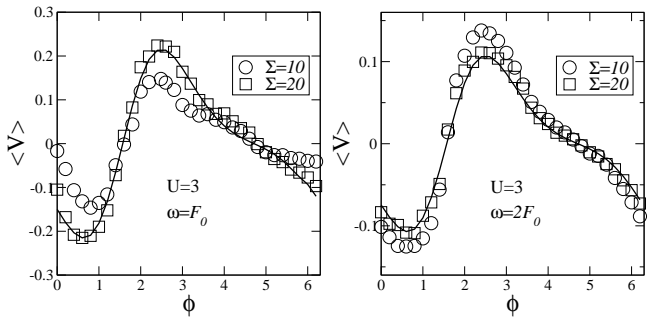

\begin{center}
\includegraphics*[clip,scale=0.27]{Fig2a.eps}
\includegraphics*[clip,scale=0.27]{Fig2b.eps}
\caption{ Phase dependence of the one-particle centroid velocity for $U=3$, two distinct initial wavepackets, $\omega=F_0$ (left panel) and $\omega=2F_0$ (right panel). The solid lines correspond to the semiclassical dependence $v\propto \cos{(\frac{F_{\omega}}{F_0}\cos(\phi)-\phi)}$ to which the results converge as $\Sigma\rightarrow\infty$. Notice the reversed dependence of the maximum velocity with $\Sigma$ at the single particle and correlated particles resonances.}
\label{fig2}
\end{center}
\end{figure}

Although the phase dependence of the drift velocity is similar to the one depicted by non-interacting particles driven by an AC field resonant with the usual BO frequency, the drift velocity at the doubled BO frequency is, ultimately, an interaction effect. In the absence of interaction ($U=0$), there is no net transport at the doubled BO resonance. Unidirectional transport of a single particle can only be promoted when the frequency of the AC field is the fundamental BO frequency (or its sub-multiples)\cite{sbo4,caetanolyra}. 

Considering that the actual value of the inter-particle interaction can be tuned via the depth of optical lattices\cite{preiss}, we computed the extremal values of the centroid velocity as a function of $U$ for both resonant cases (see fig.~3).  The  numerical result at $U=0$ and $\omega=F_0$ is consistent with the semiclassical prediction $v_{max}=2J_1(F_{\omega}/F_0=0.8)=0.73768...$ for non-interacting particles. When the interaction is turned on, the drift centroid velocity at the fundamental resonance initially decreases, passing at a minimum value on an intermediate coupling strength. On the other hand, a net transport develops at the doubled BO resonance, reaching a maximum also at a finite $U$. The maximum velocity reached at the correlated two-particles doubled BO resonance is of the same magnitude of the minimum drift velocity at the fundamental BO resonance. Such non-monotonous dependence of the drift centroid velocity on the coupling strength is due to two opposite effects played by the  interaction in the wavepacket dynamics. 
The initial wavepacket can be viewed as composed of components associated with bounded and unbounded states. The unbounded wavepacket components are not drifting under the action of the frequency doubled AC field. Therefore, the observed unidirectional motion of the wavepacket centroid is solely due to the drifting of the bounded wavepacket components. In these bounded states, the system behaves as a composite particle of charge 2$\varepsilon$. Therefore, while the velocity dependence on the field frequency is the same as that of a single (although composite) particle, its non-monotonic dependence on the interaction strength $U$ unveils its distinct influence on bounded and unbounded states.
For the components associated with bounded two-particle states, the interaction favors the pairing and promotes coherent hoppings. For the unbounded components, the interaction enhances the wavepacket width, thus reducing the double occupancy probability. As a result of these competing effects, optimal correlated two-particles motion occurs at an intermediate coupling strength\cite{wande1}. It has been previously reported that such competition also leads to a non-monotonic behavior of the BO amplitude\cite{wande3} and Anderson localization\cite{wande5,lee}.

\begin{figure}[!t]
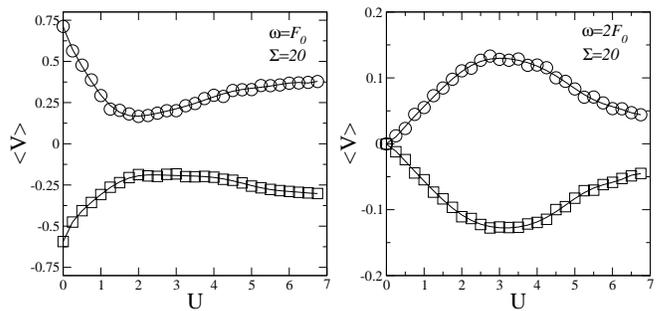

\begin{center}
\includegraphics*[clip,scale=0.25]{Fig3a.eps}
\includegraphics*[clip,scale=0.25]{Fig3b.eps}
\caption{ Maximum  one-particle drift velocity as a function of the interaction strength $U$ for an initial wavepacket with $\Sigma = 20$, $\omega=F_0$ (left panel) and $\omega=2F_0$ (right panel). The solid lines are guides to the eyes. The interaction suppresses the maximum drift velocity when the AC field is resonant with the single-particle BOs, reaching a minimum at an intermediate value of $U$. The counterpart peak in the maximum drift velocity for the resonant at the doubled BO frequency reflects the predominant role played by the two-particles bounded states in this regime.}
\label{fig3}
\end{center}
\end{figure}
 
To get more physical insight on the wavepacket dynamics at the fundamental and doubled BO resonances, we plot in fig.~4 some two-particles wavepackets after a finite evolution time. Three representative phases of the AC field were chosen, corresponding to zero, intermediate and maximal drift velocities.  The interaction strength $U=3$ favors the coherent two-particles motion. At the fundamental BO resonance, the wavepacket develops well distinct structures. The one along the diagonal stands for bounded  states with the particles located spatially close. The structure away from the diagonal stands for unbounded states with the particles driven to opposite sides of the chain. Only near the phase leading to maximal drift velocity the unbounded component is suppressed. When the AC field is resonant with the correlated two-particles doubled BO frequency, the wavepacket spreads while keeping the particles spatially close for all values of the AC field phase. Under this condition, the wavepacket is composed mainly of bounded two-particles states. This feature indicates that strong quantum correlations are always developed between the two-particles over several lattice sites. Such spatially extended quantum entangled state can be explored to probe essential quantum aspects of matter waves, such as non-locality.

\begin{figure}[!t]
\begin{center}
\includegraphics*[clip,scale=0.50]{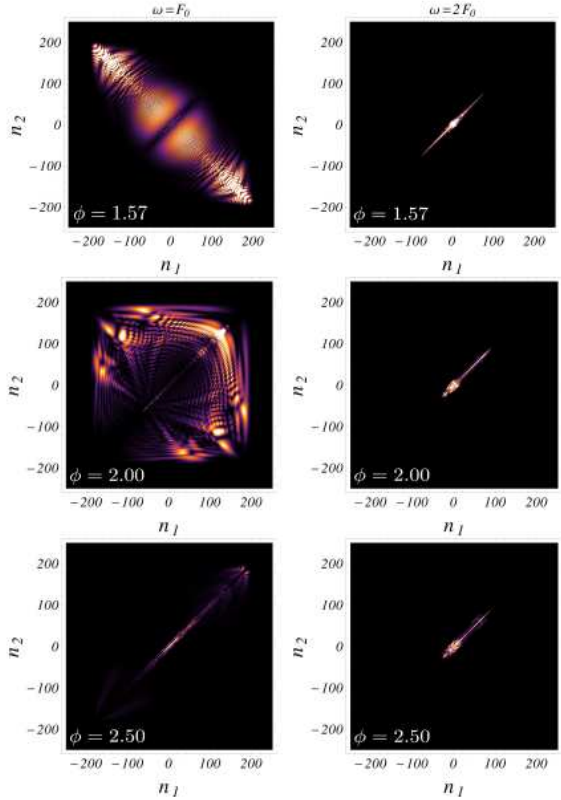}
\caption{(Color online) Density plots of wavepacket after a finite evolution time and interaction strength $U=3$. Top panels: $\phi=\pi/2$ for which the centroid velocity is zero; Intermediate panels: $\phi = 2.0$ leading to an intermediate velocity; Bottom panels: $\phi = 2.5$ resulting in maximal velocity.  Left panels: AC field resonant with the single-particle BO ($\omega=F_0$). The wavepacket develops positively correlated components (structure along the diagonal representing bounded particles) as well as anti-correlated components with the particles being driven to opposite sides (unbounded particles), except near the maximal drift velocity. Right panels:  AC field resonant with the correlated-particles BO ($\omega=2F_0$). The wavepacket remains along the diagonal (bounded particles) irrespective to the phase.}
\label{fig4}
\end{center}
\end{figure}

There are several prescriptions to quantify the degree of entanglement of a two-particles wavefunction. Here, we will use as a diagnostic tool the purity function defined as $P(t)=Tr\rho_1^2(t)$, where $\rho_1(t)$  is the reduced density matrix for particle 1 obtained after taking the partial
trace over the  states of particle 2 ($\rho_1(t) = Tr_2\rho (t)$, with $\rho(t)=|\Psi(t)\rangle\langle\Psi(t)|$). From fundamental properties of the density
matrix, the purity function $P(t) = 1$ for a pure state, meaning that the two particles are not quantum entangled. It assumes the value $P(t) = 1/N$ whenever the quantum state of particle 1 is an even
incoherent distribution among $N$ states. In the present scenario, when the two particles become maximally quantum entangled over $N$ lattice sites, i.e., $|\Psi\rangle = (1/\sqrt{N})\sum_{i=1}^N|1:n_i\rangle\otimes |2:n_i\rangle$, the partial density matrix $\rho_1 = (1/N)\sum_{i=1}^N|1:n_i\rangle\langle1:n_i|$ and the purity function reaches $P(t) = 1/N$. 
In terms of the wavefunction components, the purity function can be written as
\begin{equation} 
P = \sum_{n_1,m_1,n_2,m_2} \psi^*_{n_1,n_2}\psi_{m_1,n_2}\psi^*_{m_1,m_2}\psi_{n_1,m_2} 
\end{equation}
As a complimentary tool, we also computed the two-particles normalized correlation function defined as
\begin{equation}
C(t)=[\langle n_1n_2\rangle - \langle n_1\rangle\langle n_2\rangle]/[\langle n_1\rangle\langle n_2\rangle] .
\end{equation}

\begin{figure}[!t]
\begin{center}
\includegraphics*[clip,scale=0.33]{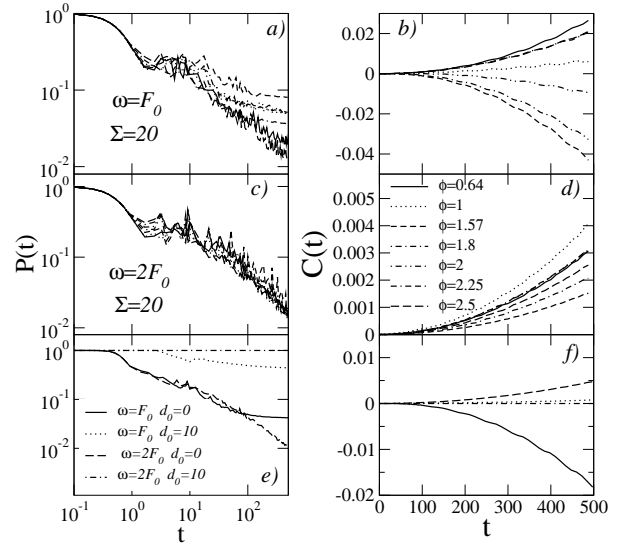}
\caption{Time evolution of the quantum purity function (left panels) and two-particles correlation function (right panels) for distinct phases of the AC field and $U=3$. (a-b): AC field  with frequency $\omega = F_0$. (c-d): AC field with frequency $\omega=2F_0$. 
At $\omega=2F_0$, the predominance of bounded states generates positive correlations and leads to an increased degree of quantum entanglement (continuously decreasing purity measure) irrespective to the phase of the AC field.
(e-f): The cases of initial Fock states ($\Sigma = 0$) with particles occupying the same $d_0=0$ and well separated $d_0=10$ sites and field phase $\phi=2.5$. Although there is no net transport, a continuously increasing degree of quantum entanglement and positive correlations is only obtained when initially close particles are driven at $\omega=2F_0$.} 
\label{fig4}
\end{center}
\end{figure}

In fig.5 we show the time evolution of the purity and pair-correlation functions in the presence of and AC field resonant with the fundamental and doubled BO frequencies.  When the particles are driven by an AC field at resonance with the fundamental BO frequency, the degree of entanglement saturates (although slower as the  extremal drift velocities are approached). Such saturation is related to the fact that unbounded states preponderate in the wavepacket dynamics (see fig. 4).   
The component with  particles driven to opposite sides of the chain has strong anti-correlations which can surpass the positive correlation associated with the bounded components. Therefore, the net pair-correlation results negative in a finite range of phase values. In contrast, the purity function continuously decrease in time when the particles are driven at resonance with the doubled BO frequency, irrespective to the phase of the AC field. This feature indicates that the wavepacket develops quantum entanglement over a continuously growing chain segment for the entire range of allowed drift velocities. The positivity of the pair-correlations reflects the predominant role played by bounded two-particles states. In Fig.5(e-f) we report results for the case of initial Fock states with the particles occupying the same site, as well as separated sites. Even depicting no net displacement of the wavepacket centroid, continuously increasing quantum entanglement and positive pair-correlations are still obtained for $\omega=2F_0$ when the particles initially occupy the same site. 

To conclude, we recall that recent experiments on cold atoms trapped in TOLs showed the capability of observing coherent atomic motion under the action of an AC field over macroscopic distances\cite{sbo3,alberti}. Further, the coherent BO of two quantum entangled particles has also been demonstrated experimentally\cite{preiss}. Therefore, the presently proposed scheme to generate and manipulate spatially extended entangled  two-atoms states by driven them using an AC field resonant with the frequency doubled two-particles BO is well within currently accessible experimental techniques. The speed and direction of the wavepacket centroid motion can be externally controlled by the AC field while the strength of the interaction can be changed by tuning the potential depth.
While the phase dependence of the drift velocity is similar to the one displayed by non-interacting particles driven by an AC field resonant with the usual BO frequency, its dependence on the interaction strength $U$ is rather non-trivial. The drift velocity vanishes in the limit of non-interacting particles and varies non-monotonically with $U$, unveiling the distinct roles played by unbounded and bounded states components on the wavepacket dynamics.
Very recently,  entanglement measures have been used to experimentally characterize the dynamics of strongly-correlated many-body systems\cite{entanglement}. Although decoherence effects due to inherent coupling with other degrees of freedom shall be carefully taken into account, the realization of spatially extended entangled two-atoms states by exploring the phenomenon of frequency doubling of BOs may impel the development of a new class of experiments aiming to search for signatures of quantum non-locality in matter waves.

This work was supported by CNPq (Conselho Nacional de Desenvolvimento Cient\'{\i}fico e Tecnol\'ogico) and FAPEAL ( Funda\c{c}\~ao de Apoio \`a Pesquisa do Estado de Alagoas).


\end{document}